\documentclass[prb,amsmath,eqsecnum]{revtex4}
\usepackage{bm}
\begin{document}

\title{Fermi and Bose pressures in statistical mechanics}
\author{Loyal Durand}
\email[Electronic mail: ]{ldurand@hep.wisc.edu}
\affiliation{
Department of Physics, University of Wisconsin-Madison \\ Madison,
WI 53706}

\begin{abstract}
I show how the Fermi and Bose pressures in quantum systems,
identified in standard discussions through the use of
thermodynamic analogies, can be derived directly in terms of the flow of momentum across a surface by using the quantum
mechanical stress tensor. In this approach, analogous to classical kinetic theory,  pressure is naturally defined locally, a point which is obvious in terms of the stress-tensor but is hidden in the usual thermodynamic approach. The two approaches are connected by an interesting application of boundary perturbation theory for quantum systems. The treatment leads to a simple interpretation of the pressure in Fermi and Bose systems in terms of the momentum flow encoded in the wave functions. I apply the methods to several problems, investigating the properties of quasi continuous systems, relations for  Fermi and Bose pressures, shape-dependent effects and anisotropies, and the treatment of particles in external fields, and note  several interesting problems for graduate courses in statistical mechanics that arise naturally in the context of these examples. 

\end{abstract}

\maketitle

\section{Introduction}
\label{sec:intro}

 The concept of pressure in quantum systems is usually introduced in equilibrium quantum statistical mechanics using thermodynamic analogies. Thus, the Helmholtz free energy $F$ is identified with the logarithm of the the canonical partition function $Z$, and the thermodynamic potential $\Omega$, with the logarithm of the  grand partition function $\mathcal{Z}$ through the relations
 \begin{equation}
 F=-kT\ln{Z}, \qquad \Omega=-kT\ln{\mathcal{Z}}.
 \end{equation}
 Here $Z$ is defined as the usual sum over the energies in the system,
 \begin{equation}
 \label{Z_def}
Z = \textrm{ Tr}\,e^{-\beta\mathcal{H}} = \sum_\alpha e^{-\beta E_\alpha}
 \end{equation}
with $\beta=1/kT$. In the case of particle systems, a separate partition function $Z_N$ can be defined for each particle number $N$. $\mathcal{Z}$ is then defined for an indefinite number  of particles by
 \begin{equation}
 \label{Zgrand_def}
 {\mathcal{Z}} = \sum_N e^{N\beta\mu}Z_N,
 \end{equation}
 where $\mu$ is the chemical potential.
The pressure is customarily determined through one of the standard thermodymanic relations
\begin{equation}
\label{pressures}
P= -\partial F/\partial V = (kT/Z)(\partial Z/\partial V)  \quad \text{ or } \quad PV= -\Omega,
\end{equation}
with $F$ evaluated at fixed temperature $T$ and particle number $N$, and $\Omega$, at fixed $T$ and $\mu$. In the second case, the average particle number and chemical potential are related by the condition $N= -\partial\Omega/\partial\mu$.

These relations can be checked in classical particle statistical mechanics by an appeal to the results of kinetic theory, and can be further motivated by an appeal to the concept of generalized forces when the energy of a system depends explicitly on external parameters such as its volume. However,  standard discussions do not show directly how the pressure relations arise in a kinetic theory-like context in quantum statistical mechanics. For a sampling of standard treatments, see Refs.~\onlinecite{tolman,landau,feynman,terhaar,kubo,huang,mohling,betts,reichl}.

The objective of this paper is to give direct derivations of the pressure in Fermi and Bose systems using ideas analogous to those in kinetic theory, specifically the relation of pressure to momentum flow and the quantum stress tensor. These are well defined ideas in quantum systems, give a ``quantum kinetic theory'' approach to pressure, and lead to a direct interpretation of the pressure in Fermi and Bose systems in terms of the momentum flow encoded in the wave functions. I will show, in fact, that pressure is naturally defined locally, a point which is obvious in the stress-tensor approach but is hidden in the usual thermodynamic approach. The two approaches are connected by an interesting application of boundary perturbation theory for quantum systems. 

The basic ideas and relations in my approach are developed in Sec.~\ref{sec:pressure}, and their use for systems of noninteracting fermions and bosons, in Sec.~\ref{sec:applications}. I then consider several examples including the properties of quasi continuous systems, relations for  Fermi and Bose pressures, and shape-dependent effects and anisotropies in Sec.~\ref{subsec:continuous_sys}, and examples for particles in external fields in Sec.~\ref{subsec:external_fields}. The results lead to several interesting problems for graduate courses in statistical mechanics.


\section{Pressure in statistical systems}
\label{sec:pressure}

\subsection{Pressure and the stress tensor}
\label{subsec:stress_tensor}

The pressure of a system at a point $\bm{x}$ on a surface $S$ can be defined as the rate of momentum flow $d{\bm p}/dt$ across a surface element $d\bm S=\hat{n}\,dS$ at $\bm x$,
\begin{equation}
\label{P=dp/dt}
P(\bm{x},t) = \frac{d\bm p}{dt}(\bm{x},t)\cdot \hat{n},
\end{equation}
that is, in terms of the force per unit area or stress acting across the surface. $P$ depends implicitly on the orientation of the surface through $\hat{n}$, but I will not indicate this explicitly both for notational simplicity, and because the apparent $\hat{n}$ dependence is, in fact, absent for the familiar extensive, quasi continuous systems.  I will apply this definition in the quantum context. For definiteness, consider a quantum system of $N$ identical particles with the Lagrangian density 
\begin{equation}
\label{Lagrangian}
{\mathcal{L}} = \frac{i\hbar}{2}\left(\psi^*\partial_t\psi - \partial_t\psi^*\,\psi\right) - \frac{\hbar^2}{2m} \sum_{n=1}^N  (\bm{\nabla}_n \psi^*)\cdot(\bm{\nabla}_n\psi) - \psi^*\textrm{V}\psi,
\end{equation}
where $\psi=\psi(\bm{x}_1,\ldots,\bm{x}_N,t)$ is the many-particle wave function for the system and $\textrm{V}(\bm{x}_1,\ldots,\bm{x}_N)$ is the potential. The corresponding Schr\"{o}dinger equation  is 
\begin{equation}
\label{Schrodinger}
i\hbar\partial_t\psi = - \frac{\hbar^2}{2m}\sum_{n=1}^N\bm{\nabla}_n^2\psi+\textrm{V}\psi.
\end{equation}

The momentum density at a point $\bm x$ for the system in a state $|\alpha\rangle$ with wave function $\psi_\alpha$ is  
\begin{equation}
\label{mom_density}
\bm{p}_\alpha(\bm{x},t) = \frac{\hbar}{2i}\sum_{n=1}^N\int\left(\psi_\alpha^*\bm{\nabla}_n\psi_\alpha-\bm{\nabla}_n\psi_\alpha^*\,\psi_\alpha \right) \delta^3(\bm{x}-\bm{x}_n) d^3x_1\cdots d^3x_N,
\end{equation}
a result obtained by integrating the sum of single-particle momentum operators $(\hbar/2i) \left(\psi_\alpha^*\bm{\nabla}_n\psi_\alpha-\bm{\nabla}_n\psi_\alpha^*\,\psi_\alpha \right)$ over the coordinates of the unobserved particles. The integrations are over the volume $V$ in which the system is confined. Similarly, the local number density is 
\begin{equation}
\label{n(x)_def}
n_\alpha(\bm{x},t) =\sum_{n=1}^N \int\psi_\alpha^*\psi_\alpha\, \delta^3(\bm{x}-\bm{x}_n) d^3x_1\cdots d^3x_N.
\end{equation}
After some rearrangements, the time derivative of $\bm{p}_\alpha$ can be written as
\begin{eqnarray}
\label{dp/dt}
\frac{d\bm{p}_\alpha}{dt}(\bm{x},t) &=&  \frac{\hbar}{i}\sum_{n=1}^N\int\left(\partial_t\psi_\alpha^*\bm{\nabla}_n\psi_\alpha- \bm{\nabla}_n\psi_\alpha^*\partial_t\psi_\alpha\right)   \delta^3(\bm{x}-\bm{x}_n) d^3x_1\cdots d^3x_N  \nonumber \\
&&+\frac{\hbar}{2i}\sum_{n=1}^N\int\bm{\nabla}_n\left(\psi_\alpha^*\partial_t\psi_\alpha-\partial_t\psi_\alpha^*\,\psi_\alpha\right)  \delta^3(\bm{x}-\bm{x}_n) d^3x_1\cdots d^3x_N.
\end{eqnarray}

It is convenient at this point to switch to a component labeling of $\bm{p}_\alpha$ and consider $\partial_t p_{\alpha,i}$. Using the Schr\"{o}dinger equation (\ref{Schrodinger}) to eliminate the time derivatives in (\ref{dp/dt}), splitting the double sums that appear into terms with identical and different particle labels,  and organizing the results as much as possible into a set of divergences, I find after a straightforward calculation that
\begin{eqnarray}
\frac{d p_{\alpha,i}}{dt}(\bm{x},t) &=&  -\sum_{n=1}^N \int \psi_\alpha^*\left(\nabla_{n,i}\textrm{V})\psi_\alpha \delta^3(\bm{x}-\bm{x}_n \right) d^3x_1\cdots d^3x_N \nonumber 
\\
&& -\frac{\hbar^2}{2m}\sum_{n=1}^N\sum_{k=1}^3\int\nabla_{n,k}\left(\nabla_{n,i}\psi_\alpha^*\nabla_{n,k}\psi_\alpha + \nabla_{n,k}\psi_\alpha^*\nabla_{n,i}\psi_\alpha \right)  \delta^3(\bm{x}-\bm{x}_n) d^3x_1\cdots d^3x_N \nonumber \\
\label{dp/dt_2}
&& +\frac{\hbar^2}{2m}\sum_{n=1}^N\int \nabla_{n,i}\left(\nabla_{n,k}\psi_\alpha^*\,\nabla_{n,k}\psi_\alpha\right)    \delta^3(\bm{x}-\bm{x}_n)d^3x_1\cdots d^3x_N
\\
&& +\frac{\hbar^2}{4m}\sum_{n=1}^N\int \nabla_{n,i}\left(\psi_\alpha^*\bm{\nabla}_n^2\psi_\alpha + \bm{\nabla}_n^2\psi_\alpha^*\,\psi_\alpha \right)  \delta^3(\bm{x}-\bm{x}_n) d^3x_1\cdots d^3x_N \nonumber
\\
&& +\ \textrm{surface terms}, \nonumber
\end{eqnarray}
The surface terms result from the integration of divergences $\bm{\nabla}_l\cdot(\bm{\cdot})$ in variables $\bm{x}_l$ other than  the selected variable $\bm{x}_n =\bm{x}$ using Gauss' theorem. These terms vanish for the usual boundary condition for the energy eigenstates needed below, that $\psi_\alpha=0$ for any of the coordinates on the boundary of the confining volume $V$, and will be dropped.

The first term on the right hand side of (\ref{dp/dt_2}) is just the force density $F_{\alpha,i}(\bm{x},t)$ at $\bm{x}$. The remaining terms are in the form of a divergence, and the result can be written as
\begin{equation}
\label{Tij1}
\frac{dp_{\alpha,i}}{dt}(\bm{x},t) = F_{\alpha,i}(\bm{x},t) +\nabla_k T^\alpha_{k,i} (\bm{x},t),
\end{equation}
or, in dyadic notation,
\begin{equation}
\label{Tdyadic}
\frac{d\bm{p}_\alpha}{dt}(\bm{x},t) = \bm{F}_\alpha(\bm{x},t) +\bm{\nabla}\cdot\tensor{\bm{T}}^\alpha (\bm{x},t)
\end{equation}
where $T^\alpha_{k,i}$ is the quantum stress tensor evaluated in the state $|\alpha\rangle$,
\begin{eqnarray}
T^\alpha_{k,i} &=& \frac{\partial\mathcal{L}}{\partial(\partial_k\psi^*)}\partial_i\psi^* +  \frac{\partial\mathcal{L}}{\partial(\partial_k\psi)}\partial_i\psi - \mathcal{L}\,\delta_{k,i} \nonumber
\\
&=& -\frac{\hbar^2}{2m}\sum_{n=1}^N\int\left[\frac{}{} \nabla_{n,i}\psi_\alpha^*\,\nabla_{n,k}\psi_\alpha + \nabla_{n,k}\psi_\alpha^*\,\nabla_{n,i}\psi_\alpha  \right. -\delta_{k,i} \bm{\nabla}_n\psi_\alpha^*\cdot\bm{\nabla}_n\psi_\alpha
\\
&&\left. 
\label{Tij2}
 -\frac{1}{2}\delta_{k,i} \left(\psi_\alpha^*\,\bm{\nabla}_n^2\psi_\alpha + \bm{\nabla}_n^2\psi_\alpha^*\,\psi_\alpha \right)  \right]\delta^3(\bm{x}-\bm{x}_n) d^3x_1\cdots 
 d^3x_N  . \nonumber
\end{eqnarray}

Upon integrating  (\ref{Tij1}) over a volume $V'\subseteq V$, one finds that the total momentum $\bm{p}$ in $V'$ changes both because of the bulk action of the forces, and from  the flow of momentum across the boundary surface $S'=\partial V'$,
\begin{equation}
\label{Tij3}
\frac{dp_{\alpha,i}}{dt}(\bm{x},t) =  \int_{V'} F_{\alpha,i}(\bm{x},t )\,d^3x + \int_{S'} dS_k\,T^\alpha_{k,i}(\bm{x},t)
\end{equation}
or, again in dyadic notation,
\begin{equation}
\label{dP/dt_dyadic}
\frac{d\bm{p}_\alpha}{dt}(\bm{x},t) =  \int_{V'} \bm{F}_\alpha(\bm{x},t) \,d^3x + \int_{S'} d\bm{S}\cdot\tensor{\bm{T}}^\alpha(\bm{x},t).
\end{equation}
Here $ d\bm{S}\cdot\tensor{\bm{T}}$ is just the rate of momentum flow across the surface element $d\bm{S}=\hat{n}\,dS$ \textit{into} the volume $V'$, with $\hat{n}$ the outward normal to the surface.
From (\ref{P=dp/dt}), the pressure at $\bm{x}$ is given by the momentum-flow per unit area \textit{out} of $V'$. That is,
\begin{equation}
\label{Palpha0}
P_\alpha(\bm{x},t) = -\hat{n}\cdot\tensor{\bm{T}}^\alpha(\bm{x},t)\cdot\hat{n},\quad \bm{x}\in S'.
\end{equation}

For equilibrium quantum statistical mechanics, the relevant states $|\alpha\rangle$ are stationary states, that is, energy eigenstates, with $\psi_\alpha=\psi_\alpha(\bm{x}_1,\ldots,\bm{x}_N)e^{-iE_\alpha t/\hbar}$. In this case, the explicit time dependence drops out in the expressions (\ref{dp/dt_2})-(\ref{Palpha1}), and $\psi_\alpha$ can be taken in these and following expressions as the spatial wave function $\psi_\alpha(\bm{x}_1,\ldots,\bm{x}_N)$. The pressure, stress tensor, and force density are then independent of $t$, $\bm{P}_\alpha(\bm{x},t)\rightarrow\bm{P}_\alpha(\bm{x})$, $\tensor{\bm{T}}^\alpha(\bm{x},t)\rightarrow\tensor{\bm{T}}^\alpha(\bm{x})$, and  $\bm{F}_\alpha(\bm{x},t)\rightarrow\bm{F}_\alpha(\bm{x})$. Furthermore, $d\bm{p}_\alpha/dt=0$ so $-\bm{\nabla}\cdot\tensor{\bm{T}}=\bm{F}$, and the divergence of the local stress is balanced by the force density. I will specialize to this case for the remainder of the paper and use the definition
\begin{equation}
\label{Palpha1}
P_\alpha(\bm{x}) = -\hat{n}\cdot\tensor{\bm{T}}^\alpha(\bm{x})\cdot\hat{n},\quad \bm{x}\in S'.
\end{equation}
for the pressure at $\bm{x}$ in the state $|\alpha\rangle$, where $\tensor{\bm{T}}^\alpha(\bm{x})$ given by (\ref{Tij2}) with $\psi_\alpha$ the spatial wave function.

I will first consider the case in which $V'=V$ is the volume in which the system is confined, and will consider a more general case in Sec.~\ref{sec:applications}. For $\bm{x}$ on the boundary surface $S=\partial V$, $\psi_\alpha$ and the derivatives of $\psi$ parallel to the surface vanish.Thus, using (\ref{Tij2}),
\begin{equation}
\label{Palpha2}
P_\alpha(\bm{x}) = \frac{\hbar^2}{2m}\sum_{l=1}^N\int (\hat{n}\cdot\bm{\nabla}_l \psi_\alpha^* )(\hat{n}\cdot\bm{\nabla}_l \psi_\alpha)  \delta^3(\bm{x}-\bm{x}_l) d^3x_1\cdots d^3x_N, 
\end{equation}
and $P_\alpha(\bm{x})$ depends only on the normal derivatives of $\psi_\alpha$ at $\bm{x}$. Finally, weighting $P_\alpha$ by the statistical factor $e^{-\beta E_\alpha}$ and averaging over all energy eigenstates $|\alpha\rangle$,
\begin{equation}
\label{Psum}
P(\bm{x}) = \frac{1}{Z}\,\frac{\hbar^2}{2m}\sum_\alpha\sum_{l=1}^N\int (\hat{n}\cdot\bm{\nabla}_l \psi_\alpha^* )(\hat{n}\cdot\bm{\nabla}_l \psi_\alpha)e^{-\beta E_\alpha} \, \delta^3(\bm{x}-\bm{x}_l) d^3x_1\cdots d^3x_N ,
\end{equation}
where $Z$ is the canonical partition function in (\ref{Z_def}).  The sums are over all completely symmetric states for Bose systems, and over all completely antisymmetric states for Fermi systems. 

I would emphasize that the wave functions $\psi(\bm{x})_1,\ldots,\bm{x}_N$ should include any factors such as spin eigenfunctions necessary to describe internal structure that does not affect the original Schr\"{o}dinger equation. The sums over eigenstates include sums over the extra quantum numbers necessary to label the states completely. In the case of observables such as the total number density or pressure that do not depend on the internal structure, the $\psi$'s can be reduced to spatial wave functions, and the right hand sides of (\ref{Psum})  multiplied by the appropriate degeneracy factor $g$. I will follow this convention throughout the paper.

Equation(\ref{Psum}),  or the more general form in (\ref{Palpha1}), gives my basic ``quantum kinetic theory'' result. The differences between fermions and boson enter only through the symmetry properties of the wave functions and the resulting differences in the sums over states. Before going on to investigate these in simple cases, I would reemphasize that the kinetic definition of the pressure is intrinsically local and is expressed through the action of the momentum operators $-i\hbar\bm{\nabla}$ as would be expected on the basis of classical kinetic theory. It is not immediately clear how this definition of the pressure is connected with the usual ``thermodynamic'' definition in (\ref{pressures}). I will first show that the two definitions are equivalent when one considers\textit{ local} variations of the volume in the relation $P=kT\,\partial\ln{Z}/\partial V$.


\subsection{Pressure from the partition function}
\label{subsec:pressure_from_Z}

The thermodynamic definition (\ref{pressures}) of pressure in terms of the canonical partition function gives the relation
\begin{equation}
\label{deltaEsum}
P = \frac{kT}{Z}\frac{\partial Z}{\partial V} = -\frac{1}{Z}\sum_\alpha \frac{\partial E_\alpha}{\partial V} e^{-\beta E_\alpha}.
\end{equation}
A comparison of this expression with (\ref{Psum}) suggests that $\partial E_\alpha/\partial V$ should be expressible  for local variations in $V$ in terms of the normal derivative of $\psi_\alpha$ on the boundary surface. This is, in fact, easy to show using boundary perturbation theory. I consider a small change in the volume of the system implemented by moving the boundary surface outward  over a small surface patch $\Delta S$ through a normal displacement $\delta\bm{x}=\hat{n}\,\delta x(\bm{x})$ that varies smoothly over $dS$ and vanishes elsewhere. The energy $E_\alpha'$ of the system in the distorted volume will differ from the energy $E_\alpha$ of the original system, with $E'_\alpha=E_\alpha+\delta E_\alpha$. The perturbed spatial wave function $\psi'_\alpha$ and the original wave function $\psi_\alpha$ satisfy the time-independent  versions of the Schr\"{o}dinger equations  (\ref{Schrodinger}),
\begin{equation}
\label{Schrodinger2}
E'_\alpha\psi'_\alpha= -\frac{\hbar^2}{2m}\sum_{l=1}^N\bm{\nabla}_l^2\psi'_\alpha + \textrm{V}\psi'_\alpha, \qquad 
E_\alpha\psi_\alpha= - \frac{\hbar^2}{2m}\sum_{l=1}^N\bm{\nabla}_l^2\psi + \textrm{V}\psi. 
\end{equation}
Multiplying the adjoint of the first equation on the right by $\psi_\alpha$ and the second equation on the left by $\psi^{'\,*}_\alpha$ and subtracting gives the expression
\begin{equation}
\label{deltaE}
\delta E_\alpha\, \psi^{'*}_\alpha\psi_\alpha = -\frac{\hbar^2}{2m}\sum_{l=1}^N \bm{\nabla}_l \cdot \left( \bm{\nabla}_l\psi^{'*}_\alpha\, \psi_\alpha - \psi^{'*}_\alpha \,\bm{\nabla}_l\psi_\alpha\right),
\end{equation}
where I have assumed that the potential $\textrm{V}$ is unchanged. An integration of all the coordinates over the original volume gives
\begin{equation}
\label{deltaE2}
\delta E_\alpha\int_V\psi^{'*}_\alpha\psi_\alpha d^3x_1\cdots d^3x_N = -\frac{\hbar^2}{2m} \sum_{l=1}^N \int_S d\bm{S}(\bm{x}) \cdot \int_V \left( \bm{\nabla}_l\psi^{'*}_\alpha\, \psi_\alpha - \psi^{'*}_\alpha \,\bm{\nabla}_l\psi_\alpha\right) \delta^3(\bm{x}-\bm{x}_l) d^3x_1\cdots d^3x_N,
\end{equation}
where the surface integration is defined in terms the variable $\bm{x}$.

The wave function $\psi_\alpha$ vanishes on the original surface $S$, so the first term in parentheses in (\ref{deltaE2}) vanishes. Similarly,  $\psi^{'*}_\alpha=0$ on the distorted surface $S'$ so the second term vanishes except on the patch $\Delta S$ where the two surfaces differ, and the surface integration reduces to the patch $\Delta S$.  For small normal displacements $\delta\bm{x}=\hat{n}\,\delta x(\bm{x})$, $\psi'_\alpha$ can be approximated to first order on $\Delta S$ using the first nonzero term in its Taylor series expansion relative to $S'$, $\psi'_\alpha\approx-\delta\bm{x}\cdot\bm{\nabla}\psi'_\alpha\approx- (\hat{n}\cdot\bm{\nabla}\psi'_\alpha)\,\delta x$, where I note that the variation of $\psi'_\alpha$ for small displacements parallel to $\Delta S$ does not contribute to first order. Finally, taking $\psi'_\alpha$ equal to $\psi_\alpha$ in  leading order and using the normalization condition for the wave function, I find a first-order expression for $\delta E_\alpha$,
\begin{equation}
\label{deltaE3}
\delta E_\alpha = -\frac{\hbar^2}{2m}\sum_{l=1}^N \int_{\Delta S} dS\,\delta x \int_V(\hat{n}\cdot\bm{\nabla}_l\psi^*_\alpha) (\hat{n}\cdot\bm{\nabla}_l\psi^*_\alpha)  \delta^3(\bm{x}-\bm{x}_l) d^3x_1\cdots d^3x_N.
\end{equation}
The integrand in the volume integral can again be taken as constant to leading order for $\bm{x}$ on $\Delta S$ . The remaining surface integral simply gives the volume change $\delta V=\int_{\Delta S}dS\,\delta x$, so 
\begin{equation}
\label{deltaE4}
\frac{\partial E_\alpha}{\partial V}(\bm{x}) = -\frac{\hbar^2}{2m}\sum_{l=1}^N \int_V(\hat{n}\cdot\bm{\nabla}_l\psi^*_\alpha) (\hat{n}\cdot\bm{\nabla}_l\psi^*_\alpha)  \delta^3(\bm{x}-\bm{x}_l) d^3x_1\cdots d^3x_N.
\end{equation}

Substitution of this expression in (\ref{deltaEsum}) reproduces (\ref{Psum}), and the kinetic and thermodynamic definitions of the local pressure $P(\bm{x})$ agree. For homogeneous isotropic systems, the factor $(\hat{n}\cdot\bm{\nabla}\psi^*_\alpha) (\hat{n}\cdot\bm{\nabla}\psi^*_\alpha)$ in the integral in (\ref{deltaE3}) has the same value at all points on $S$, the surface integral can be extended to the entire surface, and the calculation reproduces the usual $\bm{x}$-independent expression $P=kT\, \partial\ln{Z}/\partial V$. 


\subsection{Generalizations}
\label{subsec:generalizations}

\subsubsection{Electromagnetic interactions}
\label{subsubsec:em}

The results above can be generalized in various ways. For example, in the presence of electromagnetic interactions, the Lagrangian in (\ref{Lagrangian}) becomes
\label{Lagrangian_em}
\begin{equation}
\mathcal{L} = \frac{i\hbar}{2}\left(\psi^*\partial_t\psi-\partial_t\psi^*\,\psi\right)-\frac{1}{2m}\sum_{n=1}^N \left[\left(i\hbar\bm{\nabla}_n - \frac{e}{c}\bm{A}\right)\psi^*\cdot\left(-i\hbar\bm{\nabla}_n - \frac{e}{c}\bm{A}\right)\psi - \psi^*\Phi\psi\right],
\end{equation}
where $\bm{A}$ and $\Phi$ are the vector and scalar potentials, $e$ is the particle charge, and the particles are again treated as identical. The pressure of the system is related to the rate of change of the kinetic momentum ``$m\bm{v}$''$=\bm{p}-\frac{e}{c}\bm{A}$ summed over the particles. A rather lengthy calculation gives the result
\begin{equation}
\label{mdv/dt}
\sum_{n=1}^N\frac{d}{dt}\left(\bm{p}_{n,\alpha}-\frac{e}{c}\bm{A}_\alpha\right)(\bm{x},t) = \bm{F}_\alpha(\bm{x},t) +\bm{\nabla}\cdot\tensor{\bm{T}}^\alpha(\bm{x},t),
\end{equation}
where $\bm{F}$ is the Lorentz force density 
\begin{eqnarray}
\bm{F}_\alpha(\bm{x},t)& =& \sum_{n=1}^N\int e\psi^*_\alpha\bm{E}\psi_\alpha \,\delta^3(\bm{x}-\bm{x}_n)d^3x_1\cdots d^3x_N \\
\label{em_force}
&& +\frac{e}{2mc}\sum_{n=1}^N\int \left[\left(i\hbar\bm{\nabla}_n-\frac{e}{c}\bm{A}\right)\psi_\alpha^*\times\bm{B} \psi_\alpha - \psi^*_\alpha\bm{B}\times\left(i\hbar\bm{\nabla}-\frac{e}{c}\bm{A}\right)\psi_\alpha\right]\,\delta^3(\bm{x}-\bm{x}_n)d^3x_1\cdots d^3x_N,
\nonumber
\end{eqnarray}
and $T^\alpha_{k,i}$ is the gauge-invariant stress tensor,
\begin{eqnarray}
T_{k,i}^\alpha  &=&  -\frac{1}{2m}\sum_{n=1}^N\int\left[\frac{}{} \left(i\hbar \nabla_{n,i} - \frac{e}{c}A_i\right)\psi_\alpha^*\left(-i\hbar \nabla_k-\frac{e}{c}A_k\right)\psi_\alpha + \left(i\hbar \nabla_{n,k} - \frac{e}{c}A_k\right)\psi_\alpha^*\left(-i\hbar \nabla_i-\frac{e}{c}A_i\right)\psi_\alpha \right. \nonumber
\\
\label{T_em}
&& -\delta_{k,i} \left(i\hbar\bm{\nabla}_n-\frac{e}{c}\bm{A}\right)\psi_\alpha^*\cdot\left(-i\hbar\bm{\nabla}_n-\frac{e}{c}\bm{A}\right)\psi_\alpha 
\\
&&\left. 
\label{Tij4}
 +\frac{1}{2}\delta_{k,i} \left(\psi_\alpha^*\left(-i\hbar\bm{\nabla}_n-\frac{e}{c}\bm{A}\right)^2\psi_\alpha + \left(i\hbar\bm{\nabla}_n-\frac{e}{c}\bm{A}\right)^2\,\psi_\alpha^*\,\psi_\alpha \right)  \right]\delta^3(\bm{x}-\bm{x}_n) d^3x_1\cdots d^3x_N . \nonumber
\end{eqnarray}
The force density has the expected form, $\bm{F}\sim e\bm{E}+(e/c)\bm{v}\times\bm{B}$.

The pressure is given by (\ref{Palpha0}), or in the equilibrium case of static fields $\bm{A}$ and $\Phi$ and stationary states $|\alpha\rangle$, by (\ref{Palpha1}). If the latter is evaluated on the confining surface where the spatial wave function $\psi_\alpha$ vanishes, the $\bm{A}$-dependent terms in (\ref{T_em}) all drop out, and the time-independent pressure $P_\alpha(\bm{x})$ is again given by the the expression in (\ref{Palpha2}) and depends only on the normal derivatives of $\psi$ and $\psi^*$. The more general expression in (\ref{Palpha1}) can be used for a surface element in the interior of the confining volume.

A calculation similar to that in Sec.~\ref{subsec:pressure_from_Z} also reproduces (\ref{Palpha2}) for static fields and stationary systems. The $\bm{A}$-dependent terms again drop out on the confining surface, and the expression for $\delta E_\alpha$ obtained in boundary perturbation theory reduces to (\ref{deltaE3}).


\subsubsection{Wave-type equations}
\label{subsubsec:wave_eqs}

Quantized systems of bosons such as photons, mesons, or phonons in a solid satisfy wave-type equations, for example, the standard wave equation
\begin{equation}
\label{wave_eq}
\frac{1}{c^2}\partial_t^2\phi - \sum_{n=1}\bm{\nabla}_n^2 \phi + \frac{m^2c^2}{\hbar^2}\phi,
\end{equation}
with or without the extra mass term or potential $m^2(\bm{x}_1,\ldots,\bm{x}_N)$. The normalization of $\phi$ is given in terms of the covariant current density,\cite{bjorken} and reduces for positive-energy eigenstates to
\begin{equation}
\label{normalization}
\int\frac{i}{c}\left(\phi^*\partial_t\phi-\partial_t\phi^*\,\phi\right)d^3x_1 \cdots d^3x_N = \frac{2E}{\hbar c}\int \phi^*\phi\,d^3x_1 \cdots d^3x_N=1. 
\end{equation}
An appropriate Lagrangian density for such systems is
\begin{equation}
\label{waveLagrangian}
\mathcal{L}=\hbar c\left(\frac{1}{c^2}\partial_t\phi^*\,\partial_t\phi - \sum_{n=1}^N \bm{\nabla}_n\phi^*\cdot\bm{\nabla}_n\phi - \phi^*\frac{m^2c^2}{\hbar^2}\phi\right).
\end{equation}
The corresponding momentum density $\bm{p}(\bm{x},t)$ is
\begin{equation}
\label{wave_mom}
\bm{p}(\bm{x},t)=-\frac{\hbar}{c}\sum_{n=1}^N\int\left(\bm{\nabla}_n\phi^*\,\partial_t\phi + \partial_t\phi^*\,\bm{\nabla}_n\phi\right)\,\delta^3(\bm{x}-\bm{x}_n)d^3x_1\cdots d^3x_N .
\end{equation}

Calculations of $d\bm{p}/dt$ similar to those above give the same formal result for the pressure as in (\ref{Palpha1}), but with the stress tensor now given by
\begin{equation}
\label{wave_stress1}
T_{k,i} = -\hbar c\sum_{n=1}^N\int\left(\bm{\nabla}_{n,i}\phi^*\,\bm{\nabla}_{n,k}\phi + \bm{\nabla}_{n,k}\phi^*\,\bm{\nabla}_{n,i}\phi -\delta_{k,i}\mathcal{L} \right)\,\delta^3(\bm{x}-\bm{x}_n)d^3x_1\cdots d^3x_N .
\end{equation}
This can be written in an energy eigenstate  $|\alpha\rangle$ with wave function $\phi=\phi_\alpha(\bm{x}_1,\ldots,\bm{x}_N)e^{-iE_\alpha t/\hbar}$ as
\begin{eqnarray}
T_{k,i}^\alpha &=&  -\hbar c\sum_{n=1}^N\int\left[\bm{\nabla}_{n,i}\phi_\alpha^*\,\bm{\nabla}_{n,k}\phi_\alpha + \bm{\nabla}_{n,k}\phi_\alpha^*\,\bm{\nabla}_{n,i}\phi_\alpha -\delta_{k,i}\bm{\nabla}_n\phi_\alpha^* \cdot \bm{\nabla}_n\phi_\alpha\right. \nonumber
\\
\label{wave_stress2}
&& \left. -\frac{1}{2}\delta_{k,i}\left(\phi_\alpha^*\bm{\nabla}_n^2\phi_\alpha + \bm{\nabla}_n^2\phi_\alpha^*\,\phi_\alpha\right)\right]\,\delta^3(\bm{x}-\bm{x}_n)d^3x_1\cdots d^3x_N.
\end{eqnarray}
The corresponding pressure for $\bm{x}$ on a boundary surface $S=\partial V$ where $\phi\equiv 0$ is given  by 
\begin{equation}
\label{Pwave1}
P_\alpha(\bm{x}) = \hbar c\sum_{n=1}^N\int\left(\hat{n}\cdot\bm{\nabla}_n\phi_\alpha^* \right)\left(\hat{n}\cdot\bm{\nabla}_n\phi_\alpha\right)\delta^3(\bm{x}-\bm{x}_n) d^3x_1\cdots d^3x_N,
\end{equation}
and depends only on the normal derivatives of $\phi$ on the boundary. For a general surface $S'\subset V$, the full form of $T^\alpha_{k,i}$ must be used. The pressure in the canonical ensemble is just
\begin{equation}
\label{Pwave3}
P(\bm{x}) = \frac{1}{Z}\sum_\alpha P_\alpha(\bm{x})e^{-\beta E_\alpha}.
\end{equation}

It is easy to show that the same result follows from the usual thermodynamic relation in (\ref{deltaEsum}). Thus, from (\ref{wave_eq}),
\begin{equation}
\label{waveE^2}
\frac{E_\alpha^2}{\hbar^2c^2}\phi_\alpha = -\bm{\nabla}^2\phi_\alpha + \frac{m^2c^2}{\hbar^2}\phi_\alpha.
\end{equation}
An equation of the same form with a perturbed energy $E'_\alpha$ holds for the perturbed wave function $\phi_\alpha'$ which results from a local displacement of the boundary. Combining the two equations, I find as the analog of (\ref{deltaE}) that 
\begin{equation}
\label{Pwave2}
\frac{2E_\alpha\,\delta E_\alpha}{\hbar c}\,\phi^{'\,*}_\alpha\phi_\alpha = -\hbar c\sum_{l=1}^N \bm{\nabla}_l \cdot \left(\bm{\nabla}_l \phi^{'\,*}_\alpha\,\phi_\alpha - \phi^{'\,*}_\alpha \bm{\nabla}_l\phi_\alpha\right).
\end{equation}
Manipulations equivalent to those following (\ref{deltaE}) and the use of the normalization condition (\ref{normalization}) then give the result
\begin{equation}
\label{dEwave/dV}
\frac{\partial E_\alpha}{\partial V}(\bm{x}) = -\hbar c\sum_{l=1}^N \int\left(\hat{n}\cdot\bm{\nabla}_l\phi_\alpha^* \right)\left(\hat{n}\cdot\bm{\nabla}_l\phi_\alpha\right)\delta^3(\bm{x}-\bm{x}_l) d^3x_1\cdots d^3x_N
\end{equation}
for local variations of the boundary surface. Finally, the use of the thermodynamic relation (\ref{deltaEsum}) reproduces the expression for the local pressure in (\ref{Pwave3}). I would emphasize, however, that the equation $P_\alpha(\bm{x})=-\hat{n}\cdot\bm{T}_\alpha\cdot\hat{n}$ for the local pressure in a state $|\alpha\rangle$ holds more generally than (\ref{Pwave1}) and does not require that $\bm{x}$ be on the boundary surface. 


\section{ Pressure in noninteracting Bose and Fermi systems}
\label{sec:applications}

\subsection{Bose systems}
\label{subsec:Bose}

As a first example, I will consider the important case of noninteracting bosons in an external field. Interparticle interactions can be treated approximately using standard methods. See, for example, Refs.~\onlinecite{tolman,landau,feynman,terhaar,kubo,huang,mohling,betts,reichl}.  

The Hamiltonian for $N$ noninteracting subsystems or particles is a sum of $N$ identical single-particle Hamiltonians $H_1$, $H=\sum_{l=1}^N H_1(\bm{x}_l)$. The wave functions $\psi_k$ for the single-particle states $|k\rangle$ satisfy the  Schr\"{o}dinger  equations $H_1\psi_k(\bm{x_l})=E_k\psi_k(\bm{x_l})$. I will suppose that the energy eigenvalues have been ordered so that $E_1<E_2<E_3<\ldots$. The total energies are simply sums of single-particle energies $E_k$, and can be labeled by the number of particles in each single-particle eigenstate $|k\rangle$,
\begin{equation}
\label{sumE}
E_{n_1,n_2,\ldots} = n_1E_1+n_2E_2+\cdots \quad\textrm{where}\quad n_1+n_2+\cdots=N.
\end{equation}
The full wave function for $N$ bosons with $n_1$ in state $|k_1\rangle$,  $n_2$ in state $|k_2\rangle,\ldots$ is then a fully symmetric sum of product wave functions,
\begin{equation}
\label{Psi}
\psi_{n_1,n_2,\ldots}(\bm{x}_1,\ldots,\bm{x}_N) = \frac{1}{\sqrt{N!}}\sum_P\big{[}\overbrace{\psi_1\cdots\psi_1}^{n_1 \textrm{\ factors}} \overbrace{\psi_2\cdots\psi_2}^{n_2 \textrm{\ factors}}\cdots\big{]}\left(P(\bm{x}_1,\ldots,\bm{x}_N)\right),
\end{equation}
 where the sum is over all permutations $P$ of $N$ objects, and wave function factors with  $n_k=0$ are to be replaced by 1. The coordinates of the successive wave functions with $n_k\not=0$ are given in each term in the sum by the corresponding coordinates in the permutation $P$ of $\bm{x}_1,\ldots,\bm{x}_N$, as indicated. The set of $n_k$'s gives a unique labeling of the state. Their values are restricted by the condition $\sum_kn_k=N$.

The number density of particles at a point $\bm{x}$ is given for a definite state by (\ref{n(x)_def}). The total density reduces after the integrations to
\begin{equation}
\label{n(x)}
n_{n_1,n_2,\ldots}(\bm{x})=\sum_k n_k\psi^*_k(\bm{x}) \psi_k(\bm{x}), 
\end{equation}
and a final integration over $\bm{x}$ gives the total number of particles or independent subsystems $N$ since $\sum_k n_k=N$. Similarly, from (\ref{Palpha2}), the pressure on the boundary surface associated with the given state is
\begin{equation}
\label{Palpha3}
P_{n_1,n_2,\ldots}(\bm{x}) = \sum_k n_k \big(\hat{n}\cdot\bm{\nabla}\psi^*_k(\bm{x})\big)\big(\hat{n}\cdot\bm{\nabla}\psi_k(\bm{x})\big).
\end{equation}
More generally, for a surface $S'=\partial V',\ V'\subset V$,
\begin{equation}
\label{Palpha4}
P_{n_1,n_2,\ldots}(\bm{x}) = -\sum_k n_k\, \hat{n}\cdot\tensor{\bm{T}}^k\cdot\hat{n},\quad \bm{x}\in S',
\end{equation}
where the tensor $T_{ij}^k$ is given in (\ref{Tij2}).

It is difficult to work with the canonical distribution for bosons because of the restriction  $\sum_k n_k=N$. I will therefore change to the grand distribution as is usually done. Multiplying (\ref{n(x)}) by the Boltzmann factor $e^{-\beta E_{n_1,n_2,\ldots}}$ for the specified energy and by a factor $e^{\beta\mu N}$ which will be used to enforce the correct average number of particles, and summing over the $n_k$ and  $N$, I find that
\begin{eqnarray}
n(\bm{x})&=&\frac{1}{\mathcal{Z}}\sum_k \sum_{n_1,n_2,\ldots}\sum_N\delta_{n_1+n_2+\cdots,\,N}\, n_k\psi^*_k(\bm{x}) \psi_k(\bm{x})e^{\beta(\mu N-\sum_jn_jE_j)} \nonumber 
\\
\label{nsum}
&=&\frac{1}{\mathcal{Z}}\sum_k\left(\sum_{n_k}n_k \psi^*_k(\bm{x}) \psi_k(\bm{x})e^{-\beta(E_k-\mu)n_k}\right) \prod_{k'\not=k} \left(\sum_{n_{k'}}e^{-\beta(E_{k'}-\mu)n_{k'}}\right)
\\
&=&\frac{1}{\mathcal{Z}}\sum_k  \psi^*_k(\bm{x}) \psi_k(\bm{x}) \frac{e^{-\beta(E_k-\mu)}}{(1 - e^{-\beta(E_k-\mu)})^2} \prod_{k'\not=k}\left(\frac{1}{1 -e^{-\beta(E_{k'}-\mu)} }\right), \nonumber
\end{eqnarray}
where convegence of the sum requires that $E_k-\mu>0$. $\mathcal{Z}$ is the grand partition function,
\begin{equation}
\label{Zgrand}
\mathcal{Z} =  \prod_{k}\left(\frac{1}{1 -e^{-\beta(E_k-\mu)} }\right),
\end{equation}
so the result reduces to
\begin{equation}
\label{n_final}
n(\bm{x}) = \sum_k  \psi^*_k(\bm{x}) \psi_k(\bm{x}) \left(e^{\beta(E_k-\mu)}-1\right)^{-1}.
\end{equation}

This is just the result that would be expected. Each product of wave functions $\psi^*_k\psi_k$ appears with a weight which is just the average occupation number of the state $|k\rangle$ as calculated for the usual Bose distribution for noninteracting particles as in Refs.~\onlinecite{tolman,landau,feynman,terhaar,kubo,huang,mohling,betts,reichl}.

 In the presence of spin or other internal degeneracies, the $\psi$'s can be reduced to the spatial factors in the full wave functions, and after multiplying the right hand side of the equation by the appropriate degeneracy factor $g$, the sum in (\ref{n_final}) can be taken to run only over nondegenerate energies with the internal factors in the wave functions suppressed. 

The integral of $n(\bm{x})$ gives the average number of particles $N$ in the entire distribution,
\begin{equation}
\label{N_bose}
N =  \sum_k \left(e^{\beta(E_k-\mu)}-1\right)^{-1}.
\end{equation}
This gives an implicit relation for $\mu$ in agreement with the thermodynamic expression $N=kT(\partial\ln{\mathcal{Z}}/\partial \mu)$. For $T\rightarrow 0$, $\mu \rightarrow E_0-\frac{kT}{N}+\cdots$, and the particles collect in the ground state $|0\rangle$ with $n(\bm{x})=N\psi^*_0(\bm{x}) \psi_0(\bm{x})$.

A similar calculation gives the result for the pressure at $\bm{x}$, again of a form that could be anticipated from the single-particle forms of (\ref{Tij2}), (\ref{Palpha1}), and (\ref{Palpha2}),
\begin{eqnarray}
\label{P_final1}
P(\bm{x}) &=& -\sum_k\hat{n}\cdot \tensor{\bm{T}}^k\cdot\hat{n} \left(e^{\beta(E_k-\mu)}-1\right)^{-1}
\\
\label{P_final2}
&=&  \frac{\hbar^2}{2m}\sum_k\big(\hat{n}\cdot \bm{\nabla}\psi^*_k(\bm{x})\big)\big(\hat{n}\cdot \bm{\nabla}\psi_k(\bm{x})\big)  \left(e^{\beta(E_k-\mu)}-1\right)^{-1}.
\end{eqnarray}
The first form holds for $\bm{x}$ on a surface $S'$ inside the confining volume, the second form for $\bm{x}$ on the confining surface $S=\partial V$. For $T\rightarrow 0$, (\ref{P_final2}) gives $P(\bm{x}) \rightarrow (N\hbar^2/2m) \big(\hat{n}\cdot \bm{\nabla}\psi^*_0(\bm{x})\big)\big(\hat{n}\cdot \bm{\nabla}\psi_0(\bm{x})\big)$. 

It is interesting to note that the pressure does not vanish exactly for a system confined in a finite volume even at $T=0$, a result connected to the kinetic picture and the uncertainty relation. For example, for Bose particles in a one dimensional box of length $L$, the  wave function are $\psi_n=\sqrt{2/L}\sin{(n\pi x/L)}$, and (\ref{Palpha2}) gives $P_n=(N/L)(\hbar^2\pi^2n^2/mL^2)$, $n=1,2,\ldots$. All the particles collect in the ground state at $T=0$ (Bose condensation), but the pressure is nonzero for finite $L$,  vanishing as $L^{-2}$ for $L\rightarrow\infty$ at fixed particle density $N/L$ just as would be expected from the classical kinetic picture and the uncertainty relation $p\propto \hbar/L$.

The pressure in systems of noninteracting bosons satisfying the wave equation (\ref{wave_eq}) is also given formally by the expression in (\ref{P_final1}), with $\bm{T}$ now given by the single-particle form of (\ref{Pwave1}).  The analog of (\ref{P_final2}) is therefore
\begin{equation}
\label{P_final3}
P(\bm{x}) = \hbar c\sum_k \Big(\hat{n}\cdot \bm{\nabla}\phi_k^*(\bm{x})\Big) \Big(\hat{n}\cdot\bm{\nabla}\phi_k(\bm{x})\Big) \left(e^{\beta(E_k-\mu)}-1\right)^{-1}.
\end{equation}
%


\subsection{Fermi systems}
\label{subsec:Fermi}

The composite state of $N$ noninteracting fermions is specified completely by giving the number of particles $n_k$ in each completely labelled single-particle state $|k\rangle$
where $n_k= 0$ or $1$ only. The energy of the state  $|n_1,n_2,\ldots\rangle$ is just $E_{n_1,n_2,\ldots}=\sum_k n_k E_k$ as in (\ref{sumE}). The corresponding wave function is given by the completely antisymmetric sum
\begin{equation}
\label{Psi_fermi}
\psi_{n_1,n_2,\ldots}(\bm{x}_1,\ldots,\bm{x}_N) =  \frac{1}{\sqrt{N!}}\sum_P (-)^P \big{[}\overbrace{\psi_1}^{n_1}\overbrace{\psi_2 }^{n_2}\cdots\big{]}(P(\bm{x}_1,\ldots,\bm{x}_N)),\quad \sum_kn_k=N,
\end{equation}
where $(-)^P$ is the signature of the permutation $P$ of $N$ objects§. The factors of wavefunctions $\psi_k$ with $n_k=0$ are to be replaced by $1$. The coordinates of successive wave functions with $n_k=1$ are given in each term in the sum by the corresponding coordinates in the permutation $P$ of $\bm{x}_1,\ldots,\bm{x}_N$. The $n$'s are restricted by the condition indicated, that their sum be $N$.  

The number density and pressure of the particles in the specified state are given at a point $\bm{x}$ by (\ref{n(x)_def}) and (\ref{Palpha2}), respectively, and reduce after the integrations are performed to the expressions in (\ref{n(x)}) and (\ref{Palpha3}), or more generally (\ref{Palpha4}), just as in the bosonic case. The difference between the two cases is entirely in the allowed values of the $n$'s. The fermionic sum can be performed simply in the grand statistical distribution, with, for example,
\begin{eqnarray}
n(\bm{x}) &=& \frac{1}{\mathcal{Z}}\sum_k \sum_{n_1,n_2,\dots=0,1}\sum_N\delta_{_{n_1+n_2+\cdots,N}}\, n_k\psi^*_k(\bm{x})\psi_k(\bm{x})e^{\beta(\mu N-\sum_j n_jE_j)} \nonumber
\\
&=& \frac{1}{\mathcal{Z}}\sum_k \psi^*_k(\bm{x})\psi_k(\bm{x}) e^{-\beta(E_k-\mu)}\prod_{k'\not=k}\left(1+e^{-\beta(E_{k'}-\mu)}\right)  \nonumber
\\
\label{nsum_fermi}
&=& \sum_k \psi^*_k(\bm{x})\psi_k(\bm{x}) \left(e^{\beta(E_k-\mu)}+1\right)^{-1},
\end{eqnarray}
where I have used the relation, also easily derived,
\begin{equation}
\label{Zfermi}
\mathcal{Z} = \prod_k\left(1 + e^{-\beta(E_k-\mu)}\right).
\end{equation}

The expression in (\ref{nsum_fermi}) is again what would be expected since the final factor is just the average occupation number of the state $|k\rangle$ in the grand ensemble. Similarly, using (\ref{Palpha3}),
\begin{equation}
\label{Pfermi_final}
P(\bm{x}) = \frac{\hbar^2}{2m}\sum_k\Big(\hat{n}\cdot \bm{\nabla}\psi^*_k(\bm{x})\Big)\Big(\hat{n}\cdot\bm{\nabla}\psi(\bm{x})\Big)\left(e^{\beta(E_k-\mu)}+1\right)^{-1},
\end{equation}
or more generally,
\begin{equation}
\label{Pfermi_final2}
P(\bm{x}) = -\sum_k \hat{n}\cdot\tensor{\bm{T}}^k\cdot\hat{n}  \left(e^{\beta(E_k-\mu)}+1\right)^{-1},
\end{equation}
where $\tensor{\bm{T}}^k$ is the single-particle version of (\ref{Tij2}). In the case a system has internal spin-type degeneracies, the $\psi$'s in (\ref{Pfermi_final2}) can be reduced the spatial factors in the full wave functions, and the sum restricted to the nondegenerate spatial eigenstates after multiplying the right hand side of the equation by the degeneracy factor $g$,


\section{Examples}
\label{sec:examples}

\subsection{Quasi continuous systems}
\label{subsec:continuous_sys}

\subsubsection{General considerations}
\label{subsubsec:general}

The systems to which statistical descriptions are applied most frequently are large, extensive systems in which the potentials are uniform or periodic. A well-known theorem shows that the number of eigenvalues $E_k$ smaller than a fixed value $E$ grows proportionally to the volume of the system for $V\rightarrow\infty$. [See Kac, Ref.~\onlinecite{kac}, for a famous discussion of this result and its history in the context of the spectrum of a drum.] The eigenvalues therefore pack together for $V$ large, surface effects on the spectrum become negligible, and it is plausible that the sums over states in the preceding sections can be converted to integrals when there are many states with energies less than $kT$. The main question concerns the behavior of the wave functions in the limit of large $V$. We expect, in fact, that the products $\psi^*(\bm{x})\psi(\bm{x})$ and  $\bm{\nabla}\psi^*(\bm{x})\cdot\bm{\nabla}\psi(\bm{x})$  will each reduce for large $V$  to the sum of a term describing their smooth average behavior, and extra rapidly oscillating terms that average approximately to zero. The result should again be insensitive to surface effects for $V$ sufficiently large. 

These ideas can be illustrated for a uniform system in $D$ dimensions by quantizing in a box with sides $L_i$, $i=1,2,\ldots,D$. The wave functions and energies are
\begin{equation}
\label{psi_box}
\psi_{\{n\}}(\bm{x}) = \prod_{i=1}^D\sqrt{\frac{2}{L_i}}\sin{\frac{\pi n_ix_i}{L_i}}, \qquad E_{\{n\}}=\sum_{i=1}^D n_i^2\frac{h^2}{8mL_i^2},
\end{equation}
where $\{n\}$ is a multi index, $\{n\}=(n_1,\ldots,n_D)$ with $n_i=1,2,\ldots$ only. This gives
\begin{equation}
\label{psi_box2}
\psi_{\{n\}}^*(\bm{x})\psi_{\{n\}}(\bm{x}) = \frac{1}{V_D}\prod_{i=1}^D\left(1-2\cos{\frac{2\pi n_ix_i}{L_i}}\right),
\end{equation}
where $V_D=\prod_iL_i$ is the volume of the $D$-dimensional parallelepiped in which the system is confined and the state label $k$ in earlier equations is now given explicitly by the multi index $n_1,n_2,\ldots,n_D$. 

The typical index for states excited at temperature $T$ is $n_{ex}\sim (8mkTV^{2/D}/h^2)^{1/2}$. If this is large, many states will be excited as required for the conversion of sums to integrals, and the oscillating terms in (\ref{psi_box2}) will average to zero over small regions of the box. Then, for observations over such regions, $\psi^*\psi\approx 1/V_D$, a result independent of the shape of $V_D$. This is the same result as that obtained using the standard approximation of running waves with periodic boundary conditions, $\psi\approx (1/\sqrt{V})\exp{(\sum_i \pi n_ix_i/L_i)}$. An independent argument shows that the sums of the oscillating terms vanish rapidly at fixed $\bm{x}$ as the numbers of significant terms in the summations grow, that is, for many states excited. Since the level spacings tend to zero for $V_D\rightarrow\infty$, either argument shows that only the leading term in (\ref{psi_box2}) is important for spatially large systems.

Dropping the oscillating terms in (\ref{psi_box2}), the expression for the local number density for uniform Fermi and Bose systems becomes
\begin{equation}
\label{psi_box3}
n(\bm{x}) \approx \frac{1}{V_D}\sum_{n_1,\ldots,n_D} \left(e^{\beta(E_{n_1,\ldots,n_D}-\mu)}\pm 1\right)^{-1} ,
\end{equation}
where the upper and lower signs refer to Fermi and Bose systems, respectively. The sums can be converted approximately to integrals by repeated use of the Euler-Maclaurin summation formula
\begin{equation}
\label{euler-maclaurin}
\sum_{n=1}^\infty f(n) = \int_0^\infty f(n)\,dn - \frac{1}{2}f(0) - \frac{1}{12}f'(0) + \frac{1}{720}f^{'''}(0) + \cdots.
\end{equation}
The odd-order derivatives $f^{(2k+1)}$ that appear in the Euler-Maclauring formula all vanish at $n_i=0,\,\infty$ for the function in (\ref{psi_box3}). The first two terms in (\ref{euler-maclaurin}) are therefore all that survive up to exponentially small corrections that can be investigated using Poisson summation. Retaining only the leading corrections,
\begin{equation}
\label{psi_box4}
n(\bm{x}) \approx \frac{1}{V_D}\int_0^\infty dn_1dn_2\ldots dn_D  \left(e^{\beta\left(\sum_i(h^2/8mL_i^2)n_i^2-\mu\right)}\pm 1\right)^{-1} \left( 1-\frac{1}{2}\sum_i\delta(n_i)+\cdots\right).
\end{equation}

At this point, a change to the momentum variables $p_i=(h/2L_i) n_i$ gives the familiar expression for the leading term, plus corrections that vanish as $V_D^{-1/D}$ for $V_D\rightarrow\infty$,
\begin{equation}
\label{psi_box5}
n(\bm{x}) = \int \frac{d^Dp}{h^D}  \left(e^{\beta(\bm{p}_D^2/2m-\mu)}\pm 1\right)^{-1} -\sum_i \frac{h}{2L_i} \int \frac{d^{D-1}p}{h^{D-1}}  \left(e^{\beta(\bm{p}_{D-1}^2/2m-\mu)}\pm 1\right)^{-1} +O(1/L^2).
\end{equation}
The momentum integrations extend over the infinite interval $(-\infty,\infty)$, a Brillouin zone, or otherwise as appropriate. The corrections are of order $\hbar/\bar{p}_iL_i$ for $\bar{p}_i$ the typical value of the $i^{\rm{th}}$ component of the momentum in the leading term, that is, of order $\hbar/\sqrt{mkT}L_i$ for nondegenerate systems. 

Note that the final result for the leading term is isotropic in momentum space even though the original spectrum in (\ref{psi_box}) is different for motions in the different directions. This appears to be general for quasi continuous systems; see, for example, Sec.~\ref{subsubsec:WKB}. With enough energy levels occupied in the thermal distribution, the details of the spectrum become unimportant.


\subsubsection{Fermi and Bose pressures}
\label{FBpressures}

A  calculation of the pressure using the method above and either of Eqs.~(\ref{P_final2}) or (\ref{P_final3}) for Bose systems, or (\ref{Pfermi_final}) or (\ref{Pfermi_final2}) for Fermi systems  leads to analogous results for the pressure on a surface with normal $\hat{n}$,
\begin{eqnarray}
\label{FBpressure1}
P(\bm{x}) &=& \int \frac{d^Dp}{h^D}\frac{(\hat{n}\cdot\bm{p})^2}{m}  \left(e^{\beta(\bm{p}_D^2/2m-\mu)}\pm 1\right)^{-1}+\cdots 
\\
\label{FBpressure2}
& = & \frac{2}{D} \int \frac{d^Dp}{h^D} E(\bm{p}) \left(e^{\beta(\bm{p}_D^2/2m-\mu)}\pm 1\right)^{-1}+\cdots = \frac{2}{D}\langle E\rangle + \cdots,
 \end{eqnarray}
with corrections that again vanish as $V_D^{-1/D}$ for $V_D\rightarrow\infty$. The Bose and Fermi statistical factors in the integrals are isotropic in momentum space. Thus, the leading term in the expression for the pressure is independent of the direction of $\hat{n}$, and the results can be expressed in terms of the average energies as indicated.

I would emphasize that this result for the local pressure follows directly from the definition of the pressure in terms of the stress on a surface. The factors of $\hat{n}\cdot \bm{p}$ in (\ref{FBpressure1}) arise from the momentum operators $-i\hbar\bm{\nabla}$ in $-\hat{n}\cdot \tensor{\bm{T}}\cdot\hat{n}$, and correspond directly to the momenta that appear in the elementary classical derivation of the pressure in a gas. That is, the pressure is associated with the ``beating of the particles against the wall.'' 

The thermodynamic definition gives the same pressure for the quasi-homogeneous system under consideration, and the two definitions are connected by boundary perturbation theory as shown in Sec.~\ref{subsec:pressure_from_Z}. The derivation given there can be generalized to an arbitrary surface inside the volume $V$, but is only useful provided that, as here, enough is known about the wave functions to allow explicit evaluation of their derivatives.

It is easy to derive the nonclassical properties of the Fermi and Bose pressures. An integration by parts in spherical coordinates brings (\ref{FBpressure2}) to the form
\begin{equation}
\label{FBpressure3}
P(\bm{x}) =\pm kT \int \frac{d^Dp}{h^D} \ln{\left(1\pm e^{-\beta(\bm{p}_D^2-\mu)}\right)} = \mp kT \int \frac{d^Dp}{h^D} \ln{\left(1\mp \tilde{n}(\bm{p})\right)}
\end{equation}
where $ \tilde{n}(\bm{p})$ is the Fermi or Bose statistical factor in (\ref{FBpressure2}). Using the inequalities
\begin{equation}
\label{log_inequalities}
-\ln{(1- x)} >x\quad \rm{and}\quad \ln{(1+x)}<x
\end{equation}
and the fact that the integral of $\tilde{n}(\bm{p})$ gives the number density $n(\bm{x})$, one finds that 
\begin{equation}
\label{FBpressure4}
P_{\rm\,{Fermi}}(\bm{x})>n(\bm{x})kT \quad {\rm and} \quad P_{\rm{Bose}}(\bm{x})<n(\bm{x})kT.
\end{equation}
The difference clearly arises in the momentum-flow or stress picture from the necessity that the occupied single-particle states all be different for  Fermi-Dirac statistics. This forces the appearance of higher momentum states than are needed in the Bose-Einstein case, and a higher pressure for fixed $N$ and $T$.

As an example of Bose pressure, I will calculate the pressure of an equilibrium system of noninteracting neutral mesons with mass $m$. The system will be taken as extensive or quasi continuous in dimension $D$. The pressure on the boundary surface is given in a state $|\alpha\rangle$ by (\ref{Pwave1}). It can also be calculated on an interior surface using the stress tensor in (\ref{wave_stress2}) and the definition in (\ref{Palpha1}), giving the same average result for large volumes $V_D$. Using (\ref{Pwave1}) and the single-particle wave functions in a box normalized according to (\ref{normalization}), 
\begin{equation}
\label{meson_pressure1}
\phi_{p_1,\ldots,p_D}(\bm{x}) = \sqrt{\frac{\hbar c}{2E(\bm{p})}}\prod_{i=1}^D\sqrt{\frac{2}{L_i}}\sin{\frac{p_ix_i}{\hbar}}, \quad p_i=\frac{h}{2L_i}n_i, \quad n_i=1,2,\ldots,
\end{equation}
and averaging with the Bose statistical factor, I obtain
\begin{equation}
\label{meson_pressure2}
P_i(\bm{x}) = \frac{1}{V_D}\int\frac{d^Dp}{h^D}\frac{(p_i c)^2}{2E(\bm{p})}\left(e^{\beta(E(\bm{p})-\mu)}-1\right)^{-1}
\end{equation}
where $E(\bm{p})=\sqrt{\bm{p}^2c^2+m^2c^4}$. In the limit $m\rightarrow 0$, this reduces to the expression for the pressure for black body radiation or for phonons in a solid up to the necessary inclusion of the statistical factors for spins or polarizations and the use of the correct ranges of integration in the case of phonons. Thus, for black body radiation in three dimensions, including the spin degeneracy factor 2, and using the isotropy in momentum space and the fact that $\mu=0$ because photon number is not conserved,
\begin{equation}
\label{meson_pressure3}
P(\bm{x})=  \frac{1}{3V}\int\frac{d^3p}{h^3}\frac{pc}{e^{\beta pc}-1} = \frac{\hbar}{3\pi^2c^3V}\int_0^\infty d\omega\frac{\omega^3}{e^{\beta\hbar\omega}-1} = \frac{\langle E\rangle}{3V},
\end{equation}
the usual result. However, I would again emphasize that the calculation is direct, and is clearly connected to the flow of momentum across the surface through the discussion in Sec.~\ref{subsubsec:em}. No thermodynamic relations were used.

An example for Fermi systems that makes a good homework problem and shows the relation of pressure to momentum flow is the calculation of the Fermi pressure at $T=0$ staring from the stress tensor. The usual argument for completely degenerate Fermi systems shows that all energy levels up to a Fermi energy $E_F$ determined by $N$ must be occupied. Then from (\ref{Pfermi_final2}), $P(\bm{x})=-\sum_k\theta(E_F-E_k) \hat{n}\cdot\tensor{\bm{T}}^k\cdot\hat{n}$ where $\theta$ is the step function, $\theta(x)=1\,(0)$ for $x>(<)0$.  The remaining calculation is simple for continuous systems, and leads directly to the thermodynamic result without the use of any thermodynamic relations.


\subsubsection{Anisotropic pressures}
\label{subsubsec:anisotropic}

The corrections from the conversion of sums to integrals in (\ref{psi_box5}), and the corrections from finite-size effects in the spectrum,\cite{kac} are shape-dependent. This shape dependence leads for finite systems to anisotropic stresses or pressures. Consider, for example, the limit in which one of the dimensions of the rectangular box considered above, say $L_1$, becomes small while the other dimensions remain large. If the lowest energy for motions in the 1 direction is large on the scale of $kT$, $h^2/8mL_1^2\gg kT$, the sum over $n_1$ converges rapidly, and conversion of that sum to an integral with only small residual corrections is not possible. Keeping just the leading term in $n_1$ and treating the large dimensions in the continuum limit, the number density becomes
\begin{equation}
\label{aniso1}
n(\bm{x}) \rightarrow \frac{2}{L_1}\sin^2\frac{\pi x_1}{L_1} \int\frac{d^{D-1}p}{h^{D-1}} \left(e^{-\beta(\bm{p}_{D-1}^2/2m-\mu')}\pm 1\right)^{-1},
\end{equation}
where $\mu'=\mu-E_1$ with $E_1=h^2/8mL_1^2$ the ground state energy for motion in direction 1. Higher terms in the sum on $n_1$ are nominally suppressed by powers of $\exp{[-(n_1^2-1)E_1/kT]}\ll 1$, $n_1>1$, but the situation becomes more complicated and some excitation must occur for Fermi systems in which the Fermi energy for $N$ particles in $D-1$ dimensions exceeds $E_1$. I will not consider this refinement.

The leading factor in (\ref{aniso1}) is the absolute square of the normalized wave function $\psi_1(x_1)$ and integrates to unity. The momentum integral is independent of the remaining coordinates $x_2,\ldots,x_D$, so an integration of $n(\bm{x})$ over the full volume $V_D$ gives the total particle number $N$ as 
\begin{equation}
\label{aniso2}
N = V_{D-1} \int\frac{d^{D-1}p}{h^{D-1}} \left(e^{-\beta(\bm{p}_{D-1}^2/2m-\mu')}\pm 1\right)^{-1}.
\end{equation}
 The number density can therefore be written as
 \begin{equation}
 n(\bm{x}) = \frac{N}{V_{D-1}}\frac{2}{L_1}\sin^2\frac{\pi x_1}{L_1}.
 \end{equation}

The pressure on the wall of the box at $x_1=0$ can be calculated using the expressions in (\ref{P_final3}) and (\ref{Pfermi_final2}), with the result
\begin{equation}
\label{aniso3}
P_1 = \frac{h^2}{4mL_1^3}\frac{N}{V_{D-1}} = 2E_1\frac{N}{V_D}.
\end{equation}
 All particles must be in the $n_1=1$ state in $x_1$ for $kT\ll E_1$, with the effects of Fermi or Bose statistics absorbed in the integral factor in (\ref{aniso1}) and the corresponding factor in the expression for $P_1$.  The same result for $P_1$ holds on the surface $x_i=L_1$. Note that $P_1$ is independent of the coordinates $x_2,\ldots,x_D$ that specify the location on the surfaces at $x_1=0,\,L_1$ on which the pressure is observed. 

In contrast, the pressure on any of the remaining walls is
\begin{equation}
\label{aniso4}
P_j(\bm{x}) = \frac{2}{L_1}\sin^2\frac{\pi x_1}{L_1} \int\frac{d^{D-1}p}{h^{D-1}}\frac{p_j^2}{m} \left(e^{-\beta(\bm{p}_{D-1}^2/2m-\mu')}\pm 1\right)^{-1} =  \frac{2}{L_1}\sin^2\frac{\pi x_1}{L_1} \cdot \frac{2\langle E\rangle_{D-1}}{D-1}\frac{N}{V_{D-1}}, 
\end{equation}
$j>1$, where $\langle E\rangle_{D-1}$ the average single-particle excitation energy for a continuous system in $D-1$ dimensions. The pressure is independent of the coordinates $x_2,\ldots,x_D$, but is modulated with respect to $x_1$ by the factor $|\psi_1(x_1)|^2=(2/L_1)\sin^2(\pi x_1/L_1)$ which specifies how the $N$ particles are distributed with respect to $x_1$. The last factor in (\ref{aniso4}) is just the pressure in $D-1$ dimensions. When the distribution in $x_1$ is not observed, $x_1$ can be integrated out. The integration reduces $P_D$ to $P_{D-1}$, and the effective dimensionality of the system is reduced by one for $kT\ll E_1$, the limit in which the no thermal excitations in the 1 direction are possible. 

The effects are small in practice, with $E_1/k=0.2$\,K for a helium atom confined in a gap with $L_1=1$ nm. Anisotropies in the pressure would only be observable at lower temperatures. The situation is more complicated for electrons because of the effects of Fermi-Dirac statistics.


\subsection{Pressure and number density in an external field}
\label{subsec:external_fields}

\subsubsection{WKB approximation}
\label{subsubsec:WKB}

It is simple to treat the problem of otherwise noninteracting particles in a one-dimensional external potential $V(x)$ using the standard WKB approximation discussed in most texts on quantum mechanics (see, for example, Ref.~\onlinecite{schiff}).  The single-particle energies $E_k$ are determined in this approximation by the semiclassical quantization condition, that
\begin{equation}
\label{WKBenergy}
\left(k+\frac{1}{2}\right)h = 2\int_{x_1}^{x_2}p(E,x)\,dx
\end{equation}
for $E=E_k$, $k=0,\,1,\,2,\ldots$. Here $p(E,x)=\sqrt{2m[E-V(x)]}$ is the local momentum defined classsically at energy $E$, $h$ is Planck's constant, and $x_1$, $x_2$ are turning points in the classical motion where $p(E,x)=0$. The approximation can be shown to be good when there are many local wavelengths $h/p$ between the turning points, but tends to be good even for low-lying states in the spectrum. Since the typical excitation energy in statistical systems is $E\approx kT$, the WKB approximation will be valid provided $h^{-1}\int p(kT,x)\,dx \gg 1$. Treating $k$ and $E$ as continuous, the density of states $dk/dE$ implied by (\ref{WKBenergy}) is
\begin{equation}
\label{dn/dE}
\frac{dk}{dE} = \frac{1}{h}\int_{x_1}^{x_2} \sqrt{\frac{2m}{E-V(x)}}\,dx.
\end{equation}

The WKB wave functions can be written between the turning points as\cite{schiff}
\begin{eqnarray}
\label{WKBwave_func}
\psi_k(x) &\approx& \mathcal{N}_k \left(\frac{2m}{E_k-V(x)}\right)^{1/4}\cos{\left(\xi_k(x)-\frac{\pi}{4}\right)}\,dx,
\\
\label{xi(x)}
\xi_k(x) &=& \frac{1}{\hbar}\int_{x_1}^x \sqrt{2m[E_k-V(x)]}\,dx,
\end{eqnarray}
and decrease exponentially outside that region. Ignoring the small contributions from the exponential regions and replacing the square of the cosine by its average value of 1/2 for many oscillations in the region in which $E-V$ changes significantly, I find that the normalization constant  is related to the density of states by
\begin{equation}
\label{N_WKB}
\mathcal{N}_k^{\,2} = \frac{2}{h}\frac{dE}{dk}.
\end{equation}

Thus, following the discussion of quasi continuous systems above, the local number density in the system is
\begin{eqnarray}
n(x) &\approx& \sum_k\mathcal{N}_k^{\,2}\sqrt{\frac{2m}{E_k-V(x)}} \cos^2{\left(\xi(x)- \frac{\pi}{4}\right)} \left(e^{\beta (E_k-\mu)}\pm 1\right)^{-1} \nonumber
\\
\label{n(x)WKB}
&\approx& \frac{1}{h}\int dE\,\sqrt{\frac{2m}{E-V(x)}}\left(e^{\beta (E-\mu)}\pm 1\right)^{-1}
\\
&=& \int_{-\infty}^\infty \frac{dp}{h}\,\left(e^{\beta[p^2/2m-\mu+V(x)]}\pm 1\right)^{-1}, \nonumber
\end{eqnarray}
where in the last two lines I have first replaced the square of the cosine in $\psi_k^*\psi_k$ by its average value 1/2 and converted the sum over $k$ to an integral over $E$ using (\ref{N_WKB}), and then converted from $E$ to $p$ as the integration variable with  $p$ defined by the relation $p^2/2m=E-V(x)$. The replacement $\cos^2\rightarrow 1/2$ may be taken as a local averaging when there are many oscillations in the region observed. Alternatively, I note that the zeros of successive eigenfunctions interweave, so the zeros in the individual terms in $n(x)$ are washed out in the sum when many states are excited.

The result in (\ref{n(x)WKB}) is just that obtained through thermodynamic arguments\cite{landau} by dividing the system into small volumes over which $V(x)$ can be taken as constant, and then considering the equilibrium of the subsystems. The result, as here, is to replace the chemical potential $\mu$ in the corresponding expression  for free particles by $\mu-V(x)$. However, it is clear from the WKB-based derivation above that there are two key points in the quantum treatment. First, the average number $k$ of excited states must be large enough and vary smoothly enough for energies on the scale of $kT$ that the sum over states can be replaced by an integral. Second, the square of the wave function must oscillate sufficiently rapidly over regions in which $E-V(x)$ changes significantly that the replacement $\cos^2\rightarrow 1/2$ is valid in the sum in (\ref{n(x)WKB}). The second requirement is closely linked to the conditions needed for the validity of the WKB approximation, and for the replacement of the sum by an integral. It is worth noting in this connection that the wave function is defined over the entire volume in which the system is confined, and not just subvolumes as in Ref.~\onlinecite{landau}.

The expression in (\ref{n(x)WKB}) can be extended immediately to three dimensions for systems with with additive potentials, $V=\sum_iV_i(x_i)$. It can be extended to general potentials $V(\bm{x})$ in the form 
\begin{equation}
\label{n(x)WKB2}
n(\bm{x}) =  \int \frac{d^3p}{h^3}\,\left(e^{\beta[\bm{p}^2/2m-\mu+V(\bm{x})]}\pm 1\right)^{-1}
\end{equation}
using thermodynamic arguments as in Ref.~\onlinecite{landau}, or directly using functional integral methods such as those in in Refs.~\onlinecite{feynman-hibbs} and \onlinecite{brown}. It is again required that the oscillations in $\psi$ be rapid on the scale at which $E-V(\bm{x})$ changes significantly. The general result in (\ref{n(x)WKB2}) could probably also be derived in three dimensions using a WKB-like phase-integral approximation such as that investigated by Gutzwiller, Ref.~\onlinecite{gutzwiller}, but I have not attempted this.

Finally, the total number of particles in the system is given by the spatial integral of $n(\bm{x})$ over the confining volume,
\begin{equation}
\label{particle_number}
N = \int \frac{d^3x\,d^3p}{h^3}\,\left(e^{\beta[\bm{p}^2/2m-\mu+V(\bm{x})]}\pm 1\right)^{-1},
\end{equation}
giving a formula that can be used to determine $\mu$.

The pressure or stress in the external potential can be treated similarly. I will work in the interior of the total volume and use the general expression for the pressure given in (\ref{Palpha1}), and the single-particle form of the stress tensor in (\ref{Tij2}), specialized to one dimension. The key step involves the recognition that the derivatives in (\ref{Tij2}) can be taken to act only on the cosine factor in $\psi_k$, (\ref{WKBwave_func}). This gives
\begin{equation}
\label{dpsi/dx}
\frac{d\psi_k}{dx} \approx -\mathcal{N}_k\frac{1}{\hbar} \left(\frac{2m}{E_k-V(x)}\right)^{1/4}\sqrt{2m[E_k-V(x)]}\sin{\left(\xi_k(x)-\frac{\pi}{4}\right)}\,dx.
\end{equation}
The term omitted is of relative order $\lambda\frac{dV}{dx}/8\pi(E-V)$, and can be neglected in the region in which the WKB approximation is valid,\cite{schiff} namely that the change in the potential over a wavelength $\lambda=h/p$ is small on the scale of $E-V$. Furthermore, the correction term oscillates out of phase with the main tern, and interference effects can be neglected in averaging $\psi^*_k\psi_k$ A similar result holds for the second derivatives, with the neglected terms just those by which the WKB wave function fails to satisfy the exact Schr\"{o}dinger equation. 

The result of the calculation is
\begin{eqnarray}
P(x)&\approx& \sum_k\mathcal{N}_k^{\,2}\sqrt{2m[E_k-V(x)]}\left(e^{\beta(E_k-V(x)}\pm 1\right)^{-1} \nonumber
\\
\label{PWKB1}
&\approx& \frac{2}{h}\int dE \sqrt{2m[E_k-V(x)]}\left(e^{\beta(E_k-V(x)}\pm 1\right)^{-1}
\\
&=& \int_{-\infty}^\infty \frac{dp}{h}\frac{p^2}{m}\left(e^{\beta[p^2/2m-\mu+V(x)]}\pm 1\right)^{-1}. \nonumber
\end{eqnarray}

The expression  in  (\ref{PWKB1}) be generalized to more dimensions using thermodynamic arguments or functional integral methods, and should properly be stated in terms of the stress across a surface with normal $\hat{n}$,
\begin{equation}
\label{3Dstress}
-\hat{n}\cdot\tensor{\bm{T}}\cdot\hat{n} =  \int \frac{d^Dp}{h^D}\frac{(\hat{n}\cdot\bm{p})^2}{m}\left(e^{\beta[{\bm p}^2/2m-\mu+V(\bm{x})}\pm 1\right)^{-1}.
\end{equation}
 The final factor in (\ref{3Dstress}) is isotropic in $\bm p$ so the stress at a given point $\bf x$ is the same in all directions even for $V({\bm x})$ anisotropic. Thus,  $\langle \hat{n}\cdot\bm{p})^2\rangle={\langle\bm p}^2/D\rangle$,  and $P(\bm{x})=\frac{2}{D}K(\bm{x})$ where $K(\bm{x})$ is the average kinetic energy density at $\bf x$ with the average taken over the local statistical distribution. 

I turn next to two examples which illustrate the effects of external fields in interesting physical situations. Both lead to useful homework problems for graduate courses in statistical physics.


\subsubsection{Example: Particles in a linear potential}
\label{subsubsec:gravity}

For particles in a linear potential $V(z)=V_0\times(z/z_0)$ with  no potentials for the motion in the transverse directions, the motion in $z$ can be described in the WKB approximation as above, while the motion in the transverse coordinates can be described in terms of running waves with momenta ${\bf p}_\perp=(p_x,p_y)$. The single-particle energies are 
\begin{equation}
\label{grav_energies}
E_n({\bf p}_\perp) = \frac{p_\perp^2}{2m}+\frac{V_0}{z_0}\left(\frac{9 h^2z_0}{32mV_0}\right)^{1/3}\left(n-\frac{1}{4}\right)^{2/3}, \quad n=1,2,\ldots,
\end{equation}
where the second term is the WKB energy of the vertical motion. Many states of the vertical moton will be excited at the thermal energy $kT$ for $(kT/V_0)^{3/2}(8mV_0z_0^2/9\pi^2\hbar^2)^{1/2}\gg 1$, a condition always satisfied under realistic conditions for gases in a gravitational field with $V_0=mgz_0$ or electrons in a constant electric field $E_0$,  $V_0=eE_0z_0$. The sums over $k$ can be replaced by integrals over a momentum $p_z$ defined to reduce the second term in (\ref{grav_energies}) to standard form,
\begin{equation}
\label{p_z}
\frac{p_z^2}{2m}\equiv \frac{p_0^2}{2m}\left(n-\frac{1}{4}\right)^{2/3}, \quad {\rm where}\quad\frac{p_0^2}{2m}\equiv\frac{V_0}{z_0}\left(\frac{9 h^2z_0}{32mV_0}\right)^{1/3}.
\end{equation}

The corrections for Fermi or Bose statistics are unimportant for gases in a gravitational field under normal conditions. Following the development in Sec.~\ref{subsubsec:WKB} with the Fermi or Bose factors replaced by the simple Boltzmann factor $e^{-\beta(E_k-\mu)}$, one obtains the classical barometric equations 
\begin{equation}
\label{barometer}
P(z) = P(0)e^{-mgz/kT}, \quad n(z) = n(0)e^{-mgz/kT}, \quad P(0)=n(0)kT.
\end{equation}
For a system with area $A$, the number density $n(0)$ at $z=0$ is given in terms of the total number of particles $N$ by the integral
\begin{equation}
\label{number_calc}
N=\int d^3x\,n({\bf x})=n(0)\int_0^\infty d^3x\,e^{-mgz/kT} = n (0)A\frac{kT}{mg},
\end{equation}
so $n(0)=(N/A)(mg/kT)$ and $P(0)=Nmg/A$. The result is as expected. Note, as remarked above, that the pressure is isotropic at any point, $T_{x,x}=T_{y,y}=T_{z,z}$, even though the potential is not,  and isotropy was not used in the derivation. This is general for quasi continuous systems at sufficiently high excitation.

A more interesting result with respect to the gravitational field is the existence of a Bose-Einstein condensate in an ideal system at sufficiently low temperatures, an example that makes a good homework problem in a graduate course. The particle number for the Bose system is given in (\ref{N_bose}). This becomes
\begin{equation}
\label{BEC1}
N  \approx \frac{3A}{p_0^3h^2} \int d^2p_\perp \int_0^\infty dp_z\,p_z^2\left(e^{\beta[{\bm p}^2/2m-\mu]}-1\right) ^{-1}<  \frac{3A}{p_0^3h^2} \int d^2p_\perp \int_0^\infty dp_z\,p_z^2\left(e^{{\bm p}^2/2m}-1\right)^{-1}
\end{equation}
for the energy spectrum in (\ref{grav_energies}), with sums converted to integrals and $p_z$ defined  through (\ref{p_z}). The inequality follows from the convergence requirement that $E_0-\mu>0$ and the approximation in (\ref{BEC1}) that $E_0\approx 0$. 

The inequality is clearly violated for a fixed $N$ at sufficiently low temperatures or large values of $\beta$. It is then necessary to single out the ground state as this is given zero weight in the transition from a sum over states to an integral over $E$ or $\bm p$, and include its occupation number $N_0$ explicitly. $N$ is then given by
\begin{equation}
\label{BEC2}
N = N_0 +\frac{3A}{p_0^3h^2} \int d^2p_\perp \int_0^\infty dp_z\,p_z^2\left(e^{{\bm p}^2/2mkT}-1\right)^{-1}.
\end{equation}
The integrals can be evaluated by changing to spherical coordinates, and then to the variable $t = p^2/2m$ after performing the angular integration. The final integral gives a product of a generalized factorial or gamma function with a Riemann zeta function ,
\begin{equation}
\label{zeta}
\int_0^\infty dt\frac{t^{z-1}}{e^t-1}=\Gamma\left(z\right)\zeta\left(z\right).
\end{equation}
The inequality in (\ref{BEC1}) can just be satisfied for a given $N$ at a temperature $T=T_c$ determined by setting the right hand term equal to $N$, and fails at lower temperatures. The calculation gives
\begin{equation}
\label{BEC3}
N = \frac{3\pi^{3/2}A}{4p_0^3h^2}(2mkT_c)^{5/2}\zeta\left(5/2\right)
\end{equation}
corresponding for a gravitational potential to
\begin{equation}
\label{BEC4}
kT_c = \left[\frac{mg}{\zeta(5/2)}\left(\frac{h^2}{2\pi m}\right)^{3/2}\frac{N}{A}\right]^{2/5}.
\end{equation}
At lower temperatures,
\begin{equation}
\label{BEC5}
N_0=N\left[1-(T/T_c)^{5/2}\right].
\end{equation}

The power of $T/T_c$ in (\ref{BEC5}) is different from that for an ideal system with no field present.  The critical temperature  $T_c$ is also higher for fixed $N$ than the critical temperature $T^0_c$  in the absence of the gravitational field, $T_c/T^0_c=[\zeta(3/2)/\zeta(5/2)]^{2/3} \approx 1.56$, a difference attributable to the greater density of the gas near the ground. Finally, the ground state wave function is compact in $z$ with a characteristic extent $z_{\rm max}\approx E_1/mg=(81h^2/512mg)^{1/3} \approx 5.4\mu$m for helium, and the condensate ``falls to the floor.''

It is also interesting to note that the presence of a gravitational potential leads to the appearance of Bose-Einstein condensation in a two dimensional system with  $V=V_0(z/z_0)$ and free motion in a box of length $L$ in the transverse direction, with
\begin{equation}
\label{Tc-sD}
N_0=\left[1-(T/T_c)^2\right], \quad  kT_c=\frac{1}{m}\left(\frac{2p_0^3h}{\pi^3}\frac{N}{L}\right)^{1/2}. 
 \end{equation}
 There is no condensation for free motion in two dimensions. 

The calculations of the local number density and pressure in an external field are simple for dilute systems for which the Fermi and Bose statistical factors reduce in first approximation to the usual Boltzmann factor. They cannot be done exactly when quantum corrections are important, with many particles within a volume of a thermal wavelength cubed, but are similar numerically to the calculation of $N(\bm{x})$ in the following example.


\subsubsection{Example: Bosons in a harmonic trap}
\label{subsubsec:harmonic_trap}

An example of interest in connection with Bose-Einstein condensation is that of atoms confined in a harmonic trap with $V(\bm{x})=\sum_{i=1}^3\frac{1}{2}m\omega_i^2$. The total number of particles in the system is given by (\ref{particle_number}), specifically,
\begin{equation}
\label{trap1}
N = \int\frac{d^3x\,d^3p}{h^3}\left(e^{\beta\sum_i[p_i^2/2m+(m\omega_i^2/2)x_i^2]-\beta\mu}-1\right)^{-1}.
\end{equation}
The integral can be simplified by changing to the dimensionless variables $x'_i=\omega_i\sqrt{m/2kT}x_i$, $p'_i=p_i/\sqrt{2mkT}$, and then going to six-dimensional coordinates $\bm{s}=(\bm{x}',\bm{p}')$, $s^2=\sum_i(p_i^{'2}+x_i^{'2})$, and working in a spherical representation. The result is
\begin{equation}
\label{trap2}
N = \frac{1}{\pi^3}\left(\frac{kT}{\hbar}\right)^3\frac{1}{\omega_1\omega_2\omega_3}
\int_0^\infty \frac{s^5ds}{e^{s^2-\beta\mu}-1} \int d\Omega_6.
\end{equation}
Here $d\Omega_6$ is the element of solid angle in six dimensions, and $\int d\Omega_6=\pi^3$.

The remaining integral is maximized for $\mu=0$, and can be evaluated exactly in this limit by changing the integration variable from $s$ to $t=s^2$ and using (\ref{zeta}). The resulting equation determines the critical temperature $T_c$ for the onset of Bose-Einstein condensation for fixed particle number $N$,
\begin{equation}
\label{trap3}
kT_c = \hbar\left(\omega_1\omega_2\omega_3\frac{N}{\zeta(3)}\right)^{1/3}.
\end{equation}

For a spherical trap with an oscillation frequency $\nu=150$ Hz and $N=4\times 10^4$, fairly typical conditions for original experiments with Rb atoms, Refs.~\onlinecite{BECa,BECb,BECc}, this equation gives $T_c=6.77\,{\rm nK}\times N^{1/3}=232\,{\rm nK}$. Note that $\hbar\omega/k=7.20\,{\rm nK}\ll T_c$, so a large number of oscillator states are excited at $T_c$, and the use of the integral approximation to the sum over states is legitimate. 

The number of particles in excited states  for $T<T_c$, $\mu= 0$ is $N_{\rm excited}=N(T/T_c)^3$, and the number in the ground state is therefore
\begin{equation}
\label{trap4}
N_0=N\left[1-(T/T_c)^3\right],\quad T<T_c,
\end{equation}
with $N_0= 0$ for $T>T_c$. These calculations illuminate the conditions under which a real Bose-Einstein condensate can be formed in a gas, and make good homework problems.

The number density in a spherical trap follows from (\ref{trap1}),
\begin{equation}
\label{trap5}
n(r) = N_0(T)|\psi_0(r)|^2 + \frac{1}{h^3} \int d^3p\left(e^{\beta(p^2/2m+m\omega^2r^2/2)}-1\right)^{-1}
\end{equation}
for $T<T_c$, where $r=\sqrt{\bm{x}^2}$ and
\begin{equation}
\label{trap6}
\psi_0(r) = \frac{1}{\pi^{3/2}r_0^3} e^{-r^2/r_0^2}, \quad r_0=\sqrt{\hbar/m\omega}.
\end{equation}

The actual evaluation of the local number density in the trap from (\ref{trap5}) requires some numerical calculation but gives a striking illustration of the emergence of the condensate.  It is useful to scale $r$ by $r_0$, $n(r) = d^3N/d^3r$ by $N$, and $T$ by $T_c$, and change to $t=p^2/2m$ as the variable in the final momentum integration. Then, using (\ref{trap3}), (\ref{trap4}), and  (\ref{trap6}),
\begin{eqnarray}
\frac{1}{N}\frac{d^3N}{d^3(r/r_0)} &=& \frac{1}{\pi^{3/2}}
\left[1-\left(\frac{T}{T_0}\right)^3\right]e^{-r^2/r_0^2} \nonumber
\\
\label{trap7}
&& + \frac{1}{\sqrt{2}\,\pi^2\zeta(3)}\left(\frac{\hbar\omega}{kT_c}\right)^{3/2}\left(\frac{T}{T_c}\right)^{3/2}\int dt\,\sqrt{t}\left(e^{t+\frac{1}{2}(\hbar\omega/kT_c)(T_c/T)(r^2/r_0^2)}-1\right)^{-1}.
\end{eqnarray}
A very sharp ground-state peak appears in the initially rather broad in the number density as as $T$ is lowered below $T_c$ in qualitative agreement with the original experiments in \onlinecite{BECa,BECb,BECc}. This makes an an interesting comparison, and gives a real feeling for how the theory relates to observed Bose-Einstein condensates.

The pressure in the trapped system can be calculated similarly, and balances the force from the confining oscillator potentials. However, because of the long mean free path for particle intractions, it is not relevant for the expansion of the condensate when the confining interactions are suddenly removed.


\section{Comments}
\label{sec:comments}

The main objective of this paper was to give direct derivations of the pressures in Fermi and Bose systems using the relation of pressure to momentum flow and the quantum stress tensor. This ``quantum kinetic theory'' approach is simple conceptually,  and shows that the pressure is naturally defined locally, a point of interest for particles in external fields. It leads also to a direct understanding of the difference in Fermi and Bose pressures at fixed particle number and temperature in terms of the different momentum states excited, a point often argued qualitatively. A bonus of the analysis was the appearance of the simple examples of the use the use of boundary perturbation theory in quantum mechanics necessary to establish the connection of the of the usual thermodynamic arguments for particles and fields to the results obtained directly in stress-tensor approach.

I also discussed the properties of extensive, quasi-continuous systems, showed the role of excitations  high on the scale of $kT$ in obtaining isotropic pressures in intrinsically anisotropic systems, and illustrated the appearance of anisotropies and the effective reduction of the dimension of a system at low enough temperatures. Finally, I gave an explicit WKB derivation of the usual expressions for the number density and pressure of particles in an external field, and presented several examples which illustrate the use of the stress-tensor method in real physical problems. I have found these examples to make good homework problems in a graduate course on statistical mechanics.   


\begin{acknowledgments}
The author would like to thank the Aspen Center for Physics for its hospitality while parts of this paper were written. This work was supported in part by the U.S. Department of Energy under
Grant No.\ DE-FG02-95ER40896.
\end{acknowledgments}

\bibliography{StatMechBib}

\begin{thebibliography}{18}
\expandafter\ifx\csname natexlab\endcsname\relax\def\natexlab#1{#1}\fi
\expandafter\ifx\csname bibnamefont\endcsname\relax
  \def\bibnamefont#1{#1}\fi
\expandafter\ifx\csname bibfnamefont\endcsname\relax
  \def\bibfnamefont#1{#1}\fi
\expandafter\ifx\csname citenamefont\endcsname\relax
  \def\citenamefont#1{#1}\fi
\expandafter\ifx\csname url\endcsname\relax
  \def\url#1{\texttt{#1}}\fi
\expandafter\ifx\csname urlprefix\endcsname\relax\def\urlprefix{URL }\fi
\providecommand{\bibinfo}[2]{#2}
\providecommand{\eprint}[2][]{\url{#2}}

\bibitem[{\citenamefont{Betts and Turner}(1992)}]{betts}
\bibinfo{author}{\bibfnamefont{D.~S.} \bibnamefont{Betts}} \bibnamefont{and}
  \bibinfo{author}{\bibfnamefont{R.~E.} \bibnamefont{Turner}},
  \emph{\bibinfo{title}{Introductory Staistical Mechanics}}
  (\bibinfo{publisher}{Addison Wesley}, \bibinfo{address}{New York, N.Y.},
  \bibinfo{year}{1992}).

\bibitem[{\citenamefont{Feynman}(1974)}]{feynman}
\bibinfo{author}{\bibfnamefont{R.~P.} \bibnamefont{Feynman}},
  \emph{\bibinfo{title}{Statistical Mechanics}} (\bibinfo{publisher}{W. A.
  Benjamin, Inc.}, \bibinfo{address}{Reading, MA}, \bibinfo{year}{1974}).

\bibitem[{\citenamefont{Haar}(1966)}]{terhaar}
\bibinfo{author}{\bibfnamefont{D.~T.} \bibnamefont{Haar}},
  \emph{\bibinfo{title}{Elements of Thermostatistics}}
  (\bibinfo{publisher}{Holt Rinehart Winston}, \bibinfo{address}{New York,
  N.Y.}, \bibinfo{year}{1966}).

\bibitem[{\citenamefont{Huang}(1987)}]{huang}
\bibinfo{author}{\bibfnamefont{K.}~\bibnamefont{Huang}},
  \emph{\bibinfo{title}{Startistical Mechanics}} (\bibinfo{publisher}{John
  Wiley \& Sons}, \bibinfo{address}{New York, N.Y.}, \bibinfo{year}{1987}).

\bibitem[{\citenamefont{Kubo}(1978)}]{kubo}
\bibinfo{author}{\bibfnamefont{R.}~\bibnamefont{Kubo}},
  \emph{\bibinfo{title}{Statistical Mechanics}} (\bibinfo{publisher}{North
  Holland}, \bibinfo{address}{New York, N.Y.}, \bibinfo{year}{1978}).

\bibitem[{\citenamefont{Landau and Lifshitz}(1994)}]{landau}
\bibinfo{author}{\bibfnamefont{L.~D.} \bibnamefont{Landau}} \bibnamefont{and}
  \bibinfo{author}{\bibfnamefont{E.~M.} \bibnamefont{Lifshitz}},
  \emph{\bibinfo{title}{Statistical Physics}}
  (\bibinfo{publisher}{PegarmonPress}, \bibinfo{address}{Oxford, U.K.},
  \bibinfo{year}{1994}).

\bibitem[{\citenamefont{Mohling}(1982)}]{mohling}
\bibinfo{author}{\bibfnamefont{F.}~\bibnamefont{Mohling}},
  \emph{\bibinfo{title}{Statistical Mechanics}} (\bibinfo{publisher}{Publishers
  Creative Services}, \bibinfo{address}{Jamaica, N.Y.}, \bibinfo{year}{1982}).

\bibitem[{\citenamefont{Reichl}(1998)}]{reichl}
\bibinfo{author}{\bibfnamefont{L.~E.} \bibnamefont{Reichl}},
  \emph{\bibinfo{title}{A Modern Course in Statistical Physics}}
  (\bibinfo{publisher}{John Wiley \& Sons}, \bibinfo{address}{New York, N.Y.},
  \bibinfo{year}{1998}).

\bibitem[{\citenamefont{Tolman}(1962)}]{tolman}
\bibinfo{author}{\bibfnamefont{R.~C.} \bibnamefont{Tolman}},
  \emph{\bibinfo{title}{The Principles of Statistical Mechanics}}
  (\bibinfo{publisher}{Oxford University Press}, \bibinfo{address}{Oxford,
  U.K.}, \bibinfo{year}{1962}).

\bibitem[{\citenamefont{Bjorken and Drell}(1964)}]{bjorken}
\bibinfo{author}{\bibfnamefont{J.~D.} \bibnamefont{Bjorken}} \bibnamefont{and}
  \bibinfo{author}{\bibfnamefont{S.~D.} \bibnamefont{Drell}},
  \emph{\bibinfo{title}{Relativistic Quantum Mechanics}}
  (\bibinfo{publisher}{McGraw-Hill Book Co.}, \bibinfo{address}{New York,
  N.Y.}, \bibinfo{year}{1964}).

\bibitem[{\citenamefont{Kac}(1966)}]{kac}
\bibinfo{author}{\bibfnamefont{M.}~\bibnamefont{Kac}},
  \bibinfo{journal}{American Math. Monthly} \textbf{\bibinfo{volume}{73}},
  \bibinfo{pages}{1} (\bibinfo{year}{1966}).

\bibitem[{\citenamefont{Schiff}(1968)}]{schiff}
\bibinfo{author}{\bibfnamefont{L.~I.} \bibnamefont{Schiff}},
  \emph{\bibinfo{title}{Quantum Mechanics}} (\bibinfo{publisher}{McGraw-Hill
  Book Co.}, \bibinfo{address}{New York, N.Y.}, \bibinfo{year}{1968}).

\bibitem[{\citenamefont{Feynman and Hibbs}(1965)}]{feynman-hibbs}
\bibinfo{author}{\bibfnamefont{R.~P.} \bibnamefont{Feynman}} \bibnamefont{and}
  \bibinfo{author}{\bibfnamefont{A.~R.} \bibnamefont{Hibbs}},
  \emph{\bibinfo{title}{Quantum Mechanics and Path Integrals}}
  (\bibinfo{publisher}{McGraw-Hill Book Co.}, \bibinfo{address}{New York,
  N.Y.}, \bibinfo{year}{1965}).

\bibitem[{\citenamefont{Brown}(1992)}]{brown}
\bibinfo{author}{\bibfnamefont{L.}~\bibnamefont{Brown}},
  \emph{\bibinfo{title}{Quantum Field Theory}} (\bibinfo{publisher}{Cambridge
  University Press}, \bibinfo{address}{Cambridge, U.K.}, \bibinfo{year}{1992}).

\bibitem[{\citenamefont{Gutzwiller}(1967)}]{gutzwiller}
\bibinfo{author}{\bibfnamefont{M.~C.} \bibnamefont{Gutzwiller}},
  \bibinfo{journal}{J. Math. Phys.} \textbf{\bibinfo{volume}{8}},
  \bibinfo{pages}{1979} (\bibinfo{year}{1967}).

\bibitem[{\citenamefont{Anderson et~al.}(1995)}]{BECa}
\bibinfo{author}{\bibfnamefont{M.~H.} \bibnamefont{Anderson}}
  \bibnamefont{et~al.}, \bibinfo{journal}{Science}
  \textbf{\bibinfo{volume}{269}}, \bibinfo{pages}{198} (\bibinfo{year}{1995}).

\bibitem[{\citenamefont{Bradley et~al.}(1995)}]{BECb}
\bibinfo{author}{\bibfnamefont{C.~C.} \bibnamefont{Bradley}}
  \bibnamefont{et~al.}, \bibinfo{journal}{Phys. Rev. Lett.}
  \textbf{\bibinfo{volume}{75}}, \bibinfo{pages}{1687} (\bibinfo{year}{1995}).

\bibitem[{\citenamefont{Davis et~al.}(1995)}]{BECc}
\bibinfo{author}{\bibfnamefont{K.~B.} \bibnamefont{Davis}}
  \bibnamefont{et~al.}, \bibinfo{journal}{Phys. Rev. Lett.}
  \textbf{\bibinfo{volume}{75}}, \bibinfo{pages}{3969} (\bibinfo{year}{1995}).

\end{thebibliography}
\end{document}